\documentstyle[12pt]{article}

\newcommand{\EQ}{\begin{equation}}
\newcommand{\EN}{\end{equation}}

\newcommand{\bear}{\begin{eqnarray}}
\newcommand{\ear}{\end{eqnarray}}

\begin{document}

\topmargin 0pt
\oddsidemargin 5mm
\newcommand{\NP}[1]{Nucl.\ Phys.\ {\bf #1}}
\newcommand{\PL}[1]{Phys.\ Lett.\ {\bf #1}}
\newcommand{\NC}[1]{Nuovo Cimento {\bf #1}}
\newcommand{\CMP}[1]{Comm.\ Math.\ Phys.\ {\bf #1}}
\newcommand{\PR}[1]{Phys.\ Rev.\ {\bf #1}}
\newcommand{\PRL}[1]{Phys.\ Rev.\ Lett.\ {\bf #1}}
\newcommand{\MPL}[1]{Mod.\ Phys.\ Lett.\ {\bf #1}}
\newcommand{\JETP}[1]{Sov.\ Phys.\ JETP {\bf #1}}
\newcommand{\TMP}[1]{Teor.\ Mat.\ Fiz.\ {\bf #1}}
     
\renewcommand{\thefootnote}{\fnsymbol{footnote}}
     
\newpage
\setcounter{page}{0}
\begin{titlepage}     
\begin{flushright}
\end{flushright}
\vspace{0.5cm}
\begin{center}
\large{ An Intersecting Loop Model as a Solvable Super Spin Chain } \\
\vspace{1cm}
\vspace{1cm}
 {\large M. J.  Martins$^{1,2}$ ,  B. Nienhuis$^{1}$ and R. Rietman$^{1}$ \footnote{Present address: Philips Research Laboratories, Prof. Holstlann 4, 5656 Eindhoven, The Netherlands} } \\
\vspace{1cm}
\centerline{\em ${}^{1}$ Instituut voor Theoretische Fysica, 
Universiteit van Amsterdam }
\centerline{\em  Valcknierstraat 65, 1018 XE Amsterdam, The Netherlands}
\centerline{ \em and }
\centerline{\em ${}^{2}$ Departamento de F\'isica, 
Universidade Federal de S\~ao Carlos}
\centerline{\em Caixa Postal 676, 13565-905, S\~ao Carlos, Brasil}
\vspace{1.2cm}   
\end{center} 
\begin{abstract}
In this paper we investigate an integrable loop model and its connection
with a supersymmetric spin chain. The  Bethe Ansatz solution 
allows us to
study some properties of the ground state. When the loop fugacity $q$ lies
in the physical regime, we conjecture that the central charge is $c=q-1$
for $q$ integer $< 2$. Low-lying excitations are examined, supporting
a superdiffusive behavior for $q=1$. We argue that 
these systems are interesting examples
of integrable lattice models realizing $c \leq 0$ conformal field theories.
\end{abstract}
\vspace{.2cm}
\centerline{PACS numbers: 05.20-y, 0.5.50+q, 04.20.Jb }
\vspace{.2cm}
\end{titlepage}

\renewcommand{\thefootnote}{\arabic{footnote}}

In Statistical Mechanics the basic difference between an ordinary local 
model ( vertex models, spin chain systems ) and a loop model is that for
the latter the total weight configuration cannot be written as a product of
local weights. Being intrinsically non-local, loop models appears as ideal
paradigm for studying statistical properties of extended objects such as
polymers ( see e.g. refs. \cite{BE} ). 

In this letter we  investigate some critical properties of an integrable
intersecting loop model in a two dimensional square lattice \cite{BR}. The fact
that intersections between the polygons configurations are allowed makes
this loop model very interesting. In this case it is not clear at all how
to find a transformation to  a standard local model \cite{BA,SO}. The model
can be considered as a Lorentz gas of particles moving through a set of 
scatters randomly distributed in the nodes of the two dimensional 
square lattice \cite{RC,RC1,CHO2}. The scatters 
are double-sided mirrors allowing
right-angle reflections and they are placed along the diagonals of the square
lattice. The particles move along the edges of the lattice and 
arriving on a node can be scattered to the left, to the right or pass freely
in the case of absence of a scatter. We denote by $w_a$, $w_b$ and $w_c$ the
Boltzmann weights corresponding to these three possibilities, respectively.
By imposing periodic boundary conditions, each particle follows a closed
path. If for every closed loop we assign a fugacity $q$, then the partition
function $Z$ can be written as
\EQ
Z= \sum_{ scatter~ configurations } w_a^{n_a} w_b^{n_b} w_c^{n_c} 
q^{\# paths}
\EN
where $n_a$,$n_b$ and $n_c$ are the number of weights $w_a$, $w_b$ and $w_c$
appearing in a configuration, respectively. We noticed that strictly when
$w_c=0$ the closed loops configurations no longer intersect. In this limit
the partition function (1) can be seen as a graphical representation of
the critical $q^2$-state Potts model \cite{BA}.

Despite the inherent non-locality of this loop model it is still
possible to formulate a purely local condition that two different 
transfer-matrices commute for arbitrary system size. This is a sufficient
condition for integrability and it imposes a restriction 
on the manifold of possible weights 
$w_a$, $w_b$ and $w_c$ configurations. It turns out that 
the intersecting loop model is integrable 
 \cite{BR} if the  Boltzmann weights are parametrized as follows
\EQ
w_a(\mu)= 1-\mu,~~ w_b(\mu)= \mu,~~ w_c(\mu)= (1-\frac{q}{2}) \mu (1-\mu)
\EN

Clearly, the interesting  physical regime is when $q < 2$, where
all the ``probabilities'' $w_a$, $w_b$ and $w_b$ can be made positive 
for $ 0 < \mu < 1 $. Anyway, it is interesting to noticed that when $q$ is
an integer $\geq 2$ these weights reproduce 
( after a rescaling of $\mu$ ) the ones appearing in the $S$-matrix solution
of the $O(q)$ invariant non-linear sigma-model \cite{ZA}. This 
strongly suggest that  a mapping onto a ordinary local vertex model should not
be completely ruled out, at least for some values of $q$. Indeed, here we
argue that when $q \in \bf{Z} $ the intersecting loop model can be realized
in terms of a local spin chain which is invariant by the superorthosympletic
$Osp(n|2m)$ superalgebra. This is an important observation, since it will 
allow us to study the physical regime of (2) for $q$ integer $< 2$. We then
use the fact that the $Osp(n|2m)$ super spin chain is solvable by the Bethe 
Ansatz in order to find that their critical properties are governed by 
$c \leq 0$ conformal field theories.  We remark that $c \leq 0$ theories appear 
to be useful in condensed matter systems such as the 
quantum Hall effect \cite{HAL,HAL1}, disordered models \cite {DIS} 
and polymer field 
theories \cite{SA}. This means that our models could be
relevant lattice paradigms for some condensed matter applications, and
being integrable,  they might provide further physical insight as well.

Essential to our approach is to observe that the weights (2) can be derived
in the context of a standard Yang-Baxter solution for a local vertex model.
These weights are in one-to-one correspondence to the generators of a 
degenerated point of the Birman--Wenzel--Murakami algebra \cite{WA}. This 
algebra is generated by the identity $I_i$, a braid $b_i$ and a 
Temperely--Lieb operator $E_i$ acting on sites $i$ and $i+1$ 
of a quantum spin chain 
of length $L$. On the degenerate point the braid operator becomes a generator
of the symmetric group, namely
\EQ
b_i b_{i \pm1} b_i = b_{i\pm1} b_i b_{i\pm1},~~ b_i^2= I_i,~~ b_i b_j =
b_jb_i~~ \mbox{if} \hspace{0.3cm} |i-j| \geq 2
\EN
and the other set of relations closing the degenerated point of the
braid-monoid algebra  ( see e.g ref. \cite{MR1} ) are
\EQ
E_{i}E_{i\pm1}E_{i} = E_{i},~~ 
E_{i}^{2} = q E_{i},~~ 
E_{i}E_{j} = E_{j}E_{i}~~ \mbox{if} \hspace{0.3cm}   |i-j| \geq 2 
\nonumber \\
\EN
and
\EQ
b_{i}E_{i} = E_{i}b_{i} = E_{i},~~ 
E_{i}b_{i \pm 1} b_{i} = b_{i \pm 1}b_{i}E_{i \pm 1} = E_{i}E_{i \pm 1} 
\EN

It is not difficult to see that relations (3-5) can be baxterized
\cite{MR,MR1}, providing us with a rational solution of the Yang-Baxter 
equation having
precisely the weights $w_a$, $w_b$ and $w_c$. To make further progress we 
search for a representation of the algebraic relations (3-5). At least
for integer $q$, such representation can be found in terms of the invariants
of the superalgebra $Osp(n|2m)$ \cite{MR,MR1}. This superalgebra combines the
$O(n)$ symmetry and the simpletic $Sp(2m)$ algebra, and the integers $n$ and
$2m$ play the role of the number of bosonic and fermionic degrees
of freedom ( see e.g. ref. \cite{CO} ). The braid operator $b_i$ becomes the
graded permutation between the ($n+2m$) degrees of freedom which is defined
by \cite{KU}
\EQ
 b_{i} = \sum_{a,b=1}^{n+2m} (-1)^{p(a)p(b)} e_{ac} \otimes e_{bd}
\EN
where $p(a)$ is the parity distinguishing the bosonic ($p(a)=0$ for
$a=1, \cdots, n $) and the fermionic ($p(a)=1$ for $a=n+1, \cdots, n+2m$)
elements. Explicit expressions for the monoids $E_i$ have been 
recently discussed in detail in ref. \cite{MR1}. The important point which
matters here,
however, is that the fugacity $q$ is precisely the 
difference between the number of 
bosonic and fermionic degrees of freedom. More precisely, we have
\EQ
q= n-2m
\EN

The formulation of the corresponding transfer-matrix has to respect
the bosonic and the fermionic graduations \cite{KU,EK}. This is possible
by writing $T(\lambda)$ as the supertrace of an auxiliary monodromy operator,
$T(\lambda) =\displaystyle \sum_{a \in \cal{A}} (-1)^{p(a)} {\cal{T}}_{aa} $, 
where $\cal{A}$ stands for the horizontal space of $(n+2m)$ variables  of the
vertex model. As usual the monodromy matrix is composed by a product of vertex
operators ${\cal{L}}_{{\cal{A}}i}(\lambda)$ which are given by
\EQ
{\cal{L}}_{{\cal{A}}i}(\lambda) = (1-\frac{q}{2} -\lambda)b_i +
\lambda(1-\frac{q}{2} -\lambda) I_i + \lambda E_i
\EN

Performing the scale $\lambda \rightarrow \mu (1-q/2) $ in equation
(8), it is straightforward to see 
the correspondence between the weights (2) and the operators $I_i$, $b_i$ and
$E_i$. The corresponding local $Osp(n|2m)$ spin chain ${\cal{H}}$ 
is  proportional
to the logarithmic derivative of the transfer matrix around the regular point 
$\lambda =0$ where the vertex operator reduces 
to the graded permutation \cite{KU,EK}. The Hamiltonian can then be expressed
in terms of the operators $b_i$ and $E_i$ as
\EQ
{\cal H} = \pm \sum_{i=1}^{L} \left \{ b_{i}+\frac{1}{1-q/2} E_{i} \right \}
\EN
where the sign in (9) is chosen to select the antiferromagnetic regime of
the theory.
This supersymmetric Hamiltonian, with periodic boundary conditions imposed,
admits a Bethe Ansatz solution . This means that the 
eigenenergies $E(L)$ on a ring of size $L$ are parametrized in terms of
complex set of variables $\{ \lambda_j^{a} \}$, satisfying coupled non-linear
Bethe Ansatz equations. These equations are equivalent to the analyticity
of the eigenvalues of the corresponding transfer matrix and also reflect 
the underlying $Osp(n|2m)$ group symmetry. By taking into account
these properties ( analytical
Bethe Ansatz approach)  we can derive that they are given by
\EQ
\left[
\frac{\lambda^{(a)}_{j} -i\frac{\delta_{a,1}}{2\eta_{a}}}
{\lambda^{(a)}_{j} +i\frac{\delta_{a,1}}{2\eta_{a}}} 
\right]^{L} =
\prod_{b=1}^{r} \prod_{k=1,\; k \neq j}^{m_{b}}
\frac{\lambda^{(a)}_{j}-\lambda^{(b)}_{k} -i\frac{C_{a,b}}{2\eta_{a}}}
{\lambda^{(a)}_{j}-\lambda^{(b)}_{k} +i\frac{C_{a,b}}{2\eta_{a}}}, ~~ 
j=1, \dots, m_{a} ;~~ a=1, \dots, r
\EN
and the eigenenergies are parametrized by $\{ \lambda^{(1)}_j \}$
\EQ
E(L) = - \sum_{i=1}^{m_{1}} \frac{1}{[\lambda^{(1)}_{i}]^{2} + 1/4} + L
\EN
where $C_{ab}$ is the Cartan matrix, $r$ is the number of roots
and $\eta_a$ is the normalized length of the
$a$-th root of the $Osp(n|2m)$ superalgebra. We recall that the Dynkin diagrams
of the models with $n=1$ are special. Consequently a peculiar two-body
phase shift occurs for the last root $\{ \lambda_j^{2m} \}$. For an
algebraic Bethe Ansatz derivation of equations (10,11) for some classes of
$Osp(n|2m)$ models as well as for further 
technical details we refer to ref. \cite{MR1}.

We now turn to the study of the critical behaviour of the super spin
chain (9). The existence of a Bethe Ansatz solution allows us, in principle,
to calculate the eigenvalues $E(L)$ for  quite large values of $L$,
providing us with reasonable estimates of the finite size effects. 
For a conformally invariant system, the 
universality can  then be determined  by
exploiting a set of important relations satisfied by the eigenvalues
on a strip of size $L$ \cite{CA}. For example, the central charge 
$c$ is related to the ground state energy $E_0(L)$ by \cite{CA1}
\EQ
\frac{E_0(L)}{L} = e_{\infty} - \frac{\pi v_s c}{6 L^2} + O(L^{-2})
\EN
where $e_{\infty}$ is the ground state energy per particle in the 
thermodynamic limit and $v_s$ is the  sound velocity. These two parameters
can be determined exactly from the unitarity and the crossing properties (
around $\lambda=1-q/2$ )  of the transfer matrix $T(\lambda)$. In fact, these 
properties together imply that, in the thermodynamic limit,  
the largest eigenvalue $\Lambda_0(\lambda)$ of the transfer matrix satisfies
the relations
\EQ
\Lambda_0(\lambda) \Lambda_0(-\lambda) = 
[1-\lambda^2][(1-q/2)^2-\lambda^2],~~ \Lambda_0(\lambda) =
\Lambda_0(1-q/2 -\lambda)
\EN

Solving this equations with the restriction that the solution is free of
zeroes in the physical strip $0< \lambda < 1-q/2 $ and taking
its logarithmic derivative at $ \lambda=0$, we find that
\EQ
e_{\infty}= -\frac{1}{1-q/2} \left \{ \psi(\frac{1}{2} +\frac{1}{2-q} )
-\psi(\frac{1}{2-q}) +2 \ln(2) \right \} +1 
\EN
where $\psi(x)$ is the Euler function. The sound velocity measures 
how the energy scales with the low momenta. If we recall that equations (13) 
are identical
to the one we solve to find the crossing factors in a relativistic 
$S$-matrix theory, we can obtain that the appropriate  relativistic scale 
is given by
\EQ
v_s = \frac{\pi}{1-q/2}
\EN

We now have the basic ingredients to begin our analysis of the finite
size effects for the ground state energy. Let us first consider the case
when the fugacity is one . For this value of $q$, the partition
function of the intersecting loop model is trivial (constant) and therefore
$E_0(L)/L$ is exactly $e_{\infty}$ for any size $L$. In other words, 
all the finite size corrections to the ground state are null, and in particular
$c=0$. However, from the spin chain point of view, this scenario is far from
being trivial, and provides us with an important check concerning the loop
model $\leftrightarrow$ spin chain mapping. The simplest spin chain giving
us $q=1$ is the $Osp(3|2)$ model. Its spectrum is given in terms of one level
nested Bethe Ansatz and the  Bethe equations are
parametrized by two sets of variables $\{ \lambda_j^{(1)},
\lambda_j^{(2)} \} $. The ground state is characterized by a 
complex root distribution, forming ``fractional'' strings of the following
type
\EQ
\lambda_j^{(1)} = \xi_j^{(1)} \pm i/4 +O(e^{-L}),~~
\lambda_j^{(2)} = \xi_j^{(2)} \pm i/4 +O(e^{-L})
\EN

By solving numerically the corresponding Bethe Ansatz equation for some
values of $L$ and  substituting the value of 
$\{ \lambda_j^{(1)} \}$ in equation
(11), it is remarkable to see how the Bethe Ansatz roots conspire together
in order to produce the simple result $E_0(L)=-3L$ exactly. Note 
that for this model that $e_{\infty} =-3 $ (see equation (14)). Although 
similar effect
has been observed before in fine tuned anisotropic spin chains \cite{AL},
to the best of our knowledge, this is the first time that such simplification
is noted in a free-parameter (isotropic) set of Bethe Ansatz equations. This
gives confidence to investigate the super spin chains for other values of
$q <2$.

For $q \neq 1 $, the procedure described above can also be used, and
we have analysed the equations (10,11) for sizes up to $L=80$. In table 1
we show our estimates for the central charge $c$ for the $Osp(2|2)$,
$Osp(1|2)$, $Osp(1|4)$ and $Osp(1|6)$ supersymmetric spin chains, corresponding
to the values $q=0,-1,-3,$ and $-5$, respectively. In our numerical analysis
we already have considered the presence of logarithmic contributions of
$O(1/L^2 \ln(L) )$ to the finite size corrections of the ground state. We
remark that this kind of correction typically cannot be overcome by 
standard transfer matrix or Hamiltonian diagonalization due to size 
limitations. All the results lead us to the following conjecture for the
central charge behaviour for these models when $q$ is a integer $<2$
\EQ
c=q-1
\EN

This formula also reproduces the central charge in the limit $q=2$. For this
point, the weight $w_c$ vanishes and hence we expect the critical 
behaviour of the isotropic $6$-vertex model. Furthermore, the ground
state of the $Osp(1|2n)$ models ($q=1-2n$) are parametrized by real roots,
and by using an analytical method developed in ref. \cite{DW} we obtain 
$c=-2n$, in accordance with equation (17). Finally, it is interesting 
to note that formula (17) can be derived in the context  of a super
Sugawara
construction of the Virasoro algebra 
developed by Goddard et al \cite{GOD}. Following this work
the central charge at certain level $k$ is $c= k sdimG/(k+Q_G/2)$, where
sdimG and $Q_G$ is the superdimension and the value of the quadratic Casimir
of the superalgebra $G$. This latter expression applied to the $Osp(n|2m)$ 
superalgebra with $k=1$ gives 
$c= n-2m-1$, which agrees
with the formula we proposed for the central charge of
the loop model when $q=n-2m \leq 2$. 

In principle we
can use a similar approach to study the excitations via numerical or
analytical analysis of the Bethe Ansatz equations (10,11). This study is of
particular physical relevance for the Lorentz lattice gas $q=1$ model . In this
case an interesting quantity is the fractal dimension $d_f$ of the loops,
which characterizes the diffusion properties of the particles through the 
lattice \cite{RC,CHO2}. Recent numerical simulations \cite{CHO2} predicted
a superdiffusive behaviour $d_f=2$ as long as the density of mirrors is
smaller than one ($w_c \neq 0$). To lend some theoretical
support to this class of universality we have studied the finite size 
corrections for the lowest excitation present in the $q=1$ model. We find
that this excitation is of spin wave type, made by taking out one root $\lambda_j^2$ from the ground state configuration. In table 2 we present the finite
size sequences for the exponent $x=2h$, where $h$ is the conformal weight.
We see that the extrapolated value for $h$ is very small, indicating the
presence of a zero conformal weight and consequently 
predicting $d_f=2-2h=2$ whithin  reasonable precision.
We remark that better numerical data was prevented by strong logarithmic
corrections \cite{CHO2}.

A second example where the study of excitations plays an important role is
for the
supersymmetric $Osp(1|2)$ model $(q=-1)$. Its spectrum is parametrized
by  
a single non-linear equation and some low-lying states are amenable
of analytical study (real Bethe Ansatz roots). This model has $c=-2$ and
to obtain the Ramond sector,
having lowest conformal dimension $h=-1/8$, we need
to impose antiperiodic boundary condition on the even $U(1)$ sectors
of the superspin chain \cite{MAA}.
The twisted Hamiltonian still has a continuous $SU(2)$ symmetry
and the $U(1)$ charge plays
the role of a fermionic index . In principle, there is no inconsistency 
to interpret the twisted sector as an exited state, even though its energy lies
below the ground state, because the super spin chain is indeed non hermitian 
(though the eigenvalues are  real). This scenario is in remarkably similar 
to the one proposed in refs. \cite{HAL,HAL1} to explain certain properties
of edge excitations in the fractional Hall effect. Furthermore, concerning the
eigenspectrum of the primary fields, the twisted spin chain could either
be seen as a $c=-2$
theory or as $c=1$ Coulomb gas with radius of compactification $\rho=\sqrt{2}$.
We also noticed that the $Osp(2|2)$ spin chain is the prototype for
realizing both $c=-1$ and $c=2$ field theories. Again, our numerical data
supportes the presence of a  field
with dimension $h=-1/8$ in the twisted spin chain. Very recently, such 
relations between $c \leq 0$ and $c>0$ have been proposed and discussed in the
literature \cite{LU}. This suggests that the lattice models discussed
in this paper are realizations of non-unitary ($c <0$) conformal
theories that are related ( via appropriate twisting ) to $c >0$
systems. In particular our numerical result for the twisted $q=1$ model 
indicates central
charge $c=3$, 
suggesting that the continuum limit of the 
Lorentz gas  might be governed by a $N=2$ supersymmetric field theory (
see e.g. ref.\cite{MUS}) .  It
remains to be seen, however, what kind of physical information
can be obtained from this observation.

In summary, a solvable loop model has been mapped onto a super spin chain
and we have found its central charge behaviour  for integer values of
the fugacity in the physical regime. We hope that our discussions lend
support to the idea that these models appears to be the ideal lattice 
candidates for describing conformal properties of relevant condensed matter
systems.

\section*{Acknowledgements}
 It is a pleasure to thank J.de Gier for may useful discussions. This  
work was support by  FOM (Fundamental Onderzoek der Materie) 
and Fapesp ( Funda\c c\~ao
de Amparo \'a Pesquisa do Estado de S. Paulo), and was partially done in the
frame of Associate Membership programme of the International Centre for
Theoretical Physics, Trieste Italy.

\newpage
\underline{Table 1}: Finite size sequences for the extrapolation 
of the central charge for the models $Osp(2|2)$ ($q=0$), $Osp(1|2)$ ($q=-1$),
$Osp(1|4)$ ($q=-3$) and $Osp(1|6)$ ($q=-5$). 

\begin{table}
\begin{center}
\begin{tabular}{|c|c|c|c|c|} \hline
     $L$  &$Osp(2|2)$  &$Osp(1|2)$ & $Osp(1|4)$ & $Osp(1|6)$   \\ \hline\hline
16 & -1.11540 & -1.92337 & -4.08344 & -6.52566 \\ \hline
24 & -1.10037 & -1.93829 & -4.02750 & -6.25291 \\ \hline
32 & -1.09075 & -1.94574 & -4.00716 & -6.15381 \\ \hline
40 & -1.08401 & -1.95041 & -3.99753 & -6.10483 \\ \hline
48 & -1.07896 & -1.95369 &-3.99228 & -6.07658 \\ \hline
56 & -1.07500 & -1.95617 & -3.98917 & -6.05863 \\ \hline
64 & -1.07179 & -1.95812 & -3.98721 & -6.04643 \\ \hline
72 & -1.06911 & -1.95973 & -3.98593 & -6.03771 \\ \hline
80 & -1.06683 & -1.96107 & -3.98508 & -6.03125 \\ \hline
Extr. & -1.01 (1) & -1.996 (2) & -3.985 (3) & -5.997 (1) 
\\ \hline
\end{tabular}
\end{center}
\end{table}

\underline{Table 2}: Finite size sequences for the extrapolation 
of the  lowest conformal dimension $x=2h$ of the $q=1$ model.

\begin{table}
\begin{center}
\begin{tabular}{|c|c|} \hline
     $L$  & $ x=2h $    \\ \hline\hline
16 & $8.464651~10^{-2}$  \\ \hline
32 & $6.764093~10^{-2}$  \\ \hline
48 & $6.032298~10^{-2}$  \\ \hline
64 & $5.592822~10^{-2}$  \\ \hline
80 & $5.288533~10^{-2}$  \\ \hline
96 & $5.060282~10^{-2}$  \\ \hline
112 & $4.88002~10^{-2}$  \\ \hline
128 & $4.73245~10^{-2}$  \\ \hline
Extr. & $1.33~10^{-2}$  \\ \hline
\end{tabular}
\end{center}
\end{table}

\end{document}